\documentclass[%
 aps,
 prl,
twocolumn,
superscriptaddress,
frontmatterverbose, 
showpacs,preprintnumbers,
 nofootinbib,
 amsmath,
 amssymb,
floatfix,
latexsym,array,enumerate,letter,
]{revtex4-1}

\usepackage[utf8]{inputenc}
\usepackage[english]{babel}
\pdfoutput=1
\allowdisplaybreaks
\usepackage{graphicx,color}
\usepackage[colorlinks=true,citecolor=blue,linkcolor=blue,urlcolor=blue]{hyperref}
\usepackage{comment}
\usepackage{appendix}
\usepackage{caption}
\usepackage{subcaption}
\usepackage{footmisc}
\usepackage{bm}
\usepackage{dcolumn}
\usepackage{gensymb}
 \usepackage{float} 
\restylefloat{table}
\usepackage{multirow}
 \usepackage{hyperref}
 \usepackage{array}

\hypersetup{
    colorlinks,
    citecolor=black,
    filecolor=black,
    linkcolor=blue,
    urlcolor=blue
}

\newcounter{infer}[section]
\renewcommand*{\theinfer}{}

\newcolumntype{P}[1]{>{\centering\arraybackslash}p{#1}}
 \newcommand{\be}{\begin{equation}}
\newcommand{\ee}{\end{equation}}
\newcommand{\beq}{\begin{equation}}
\newcommand{\eeq}{\end{equation}}
\newcommand{\bea}{\begin{eqnarray}}
\newcommand{\eea}{\end{eqnarray}}

\begin{document}
\title{Hybrid stars may have an inverted structure}
\author{Chen Zhang}
\email{iasczhang@ust.hk}
\affiliation{The HKUST Jockey Club Institute for Advanced Study,
The Hong Kong University of Science and Technology, Clear Water Bay, Kowloon, Hong Kong, People’s Republic of China}
\author{Jing Ren}
\email{renjing@ihep.ac.cn}
\affiliation{Institute of High Energy Physics, Chinese Academy of Sciences, Beijing 100049, People’s Republic of China}

\begin{abstract}
We propose a new stellar structure of compact stars, the ``cross stars" that consist of a hadronic matter core and a quark matter crust, with an inverted structure compared to the conventional hybrid stars. This distinct stellar structure naturally arises from the quark matter to hadronic matter transition associated with the chemical potential crossing,
in the context of the quark matter hypothesis that either strange or up-down quark matter is the ground state of baryonic matter at low pressure. 
We find that the interplay between the hadronic matter and quark matter compositions of cross stars can help to reconcile the small radii constraints indicated by the LIGO/Virgo GW170817 event, the large radii constraints set for massive compact stars by recent NICER X-ray observations, and  the recent observation of the most-massive pulsar PSR J0952-0607. This leaves more space open for the equation of states of both hadronic matter and quark matter.

\end{abstract}
\maketitle

\section{Introduction}
Recent detections of gravitational wave (GW) signals from compact binary mergers by the LIGO/Virgo collaborations~\cite{LIGOScientific:2016aoc,  LIGOScientific:2017bnn,LIGOScientific:2018mvr,TheLIGOScientific:2017qsa,Abbott:2018wiz,Abbott:2020uma,Abbott:2020khf,LIGOScientific:2018cki} have greatly moved our understanding of black holes and compact stars forward. In particular, they provide unprecedented tools to probe the QCD matter phases, such as hadronic matter and quark matter.  

Hadronic matter (HM) is conventionally assumed to be the ground state of baryonic matter at zero temperature and pressure. Quark matter (QM)
then turns energetically favorable only in extreme environments such as neutron star interiors.
This leads to the conventional hybrid stars composed of a QM core and HM crust~\cite{Alford:2013aca,Alford:2004pf}. Some models of hybrid stars have been found to be helpful to relieve the tension among different astrophysical observations~\cite{Dexheimer:2018dlz,Li:2021sxb,Most:2019onn, Li:2022ivt,Huang:2022mqp,Kedia:2022nns,Li:2021crp,Li:2022cfd,Miao:2020yjk}.

On the other hand, it has long been realized that the ground state of baryonic matter at zero pressure and temperature might actually be QM, dubbed as the QM hypothesis~\cite{Bodmer:1971we,Witten,Terazawa:1979hq,Holdom:2017gdc}. The first candidate was strange quark matter (SQM) with comparable numbers of $u, \,d, \,s$ quarks~\cite{Bodmer:1971we,Witten,Terazawa:1979hq}. More recently, it was demonstrated that $u, d$ quark matter ($ud$QM) could be more stable than SQM and ordinary nuclear matter at a sufficiently large baryon number~\cite{Holdom:2017gdc},
if taking the flavor-dependent feedback of the quark gas on the QCD vacuum into account. 
The SQM and $ud$QM hypotheses  then allow the possibility of bare quark stars (QSs), such as strange quark stars (SQSs) that consist of SQM~\cite{Haensel:1986qb,Alcock:1986hz,Zhou:2017pha,Miao:2021nuq,Weber:2004kj,Kumari:2021tik,Yang:2020ykt,Lai:2022yky,Li:2022rxw,Li:2020wbw}  or up-down quark stars ($ud$QSs) that consist of $ud$QM~\cite{Zhang:2019mqb,Wang:2019jze,Zhao:2019xqy, Ren:2020tll, Cao:2020zxi, Yuan:2022dxb,Xia:2020byy, Xia:2022tvx,Li:2022vof,Wang:2021byk}.  

It's important to differentiate between the absolutely stable quark matter in this context and that inside conventional hybrid stars, which originates from the deconfinement of hadronic matter. 
Additionally, the HM equation of state  (EOS) inside compact stars is subject to larger uncertainties~\cite{Oter:2019rqp, Kurkela:2014vha}, due to the lack of knowledge at the intermediate pressure range. 
Therefore, under the QM hypothesis, it's possible for QM to transit to HM inside the quark stars if the latter becomes more stable than the former in the unexplored regions.
This gives rise to a new type of stellar objects, the \emph{Cross stars} (CrSs), due to the crossing of chemical potentials of the two phases~\cite{footnote1}.
It consists of a QM crust and a HM core, and thus features an inverted structure compared to the conventional hybrid stars.  
It is well known that pure QSs can form from neutron stars hit by QM nuggets~\cite{Farhi:1984qu,Xia:2022tvx} or quantum nucleation in the interior~\cite{Iida:1998pi,Bombaci:2016xuj,Ren:2020tll}. If HM is more stable than QM at some density, then the same process can form CrSs directly or from the quantum nucleation of HM inside QSs. The phase transition needs the center pressure larger than the one for crossing, which can happen from QS spin-down, accretion or merger.

We will demonstrate that CrSs can be naturally realized within simple theoretical frameworks, and also accommodate different astrophysical observations.

\section{Crossing from QM to HM}

To describe the phase transition from QM to HM inside CrSs, we adopt the Maxwell construction, which assumes a sharp phase transition with no mixed phase. This is a valid assumption here considering the rapid hadron-quark phase transition in the empirical range of the surface tension of the hadron-quark interface~\cite{Iida:1998pi,Bombaci:2016xuj, Ren:2020tll, Alford:2001zr,Li:2021sxb}. 
The phase transition is then determined by the crossing of chemical potentials for the two matter phases at low densities.
Given the general thermodynamic relation between the chemical potential and EOS~\cite{footnoteThermo},
the QM to HM transition occurs in the case of a softer HM EOS and stiffer QM EOS at low densities. Below, we will conduct a comprehensive analysis to determine the feasibility of this transition within the extensively studied parameter space of QM and HM. Similar transitions have also been studied before, but in the context of specific holographic models~\cite{Annala:2017tqz,Jokela:2018ers, Hoyos:2021uff}.

We start by describing the properties of QM. 
The existence of absolutely stable quark matter at low densities originates from the reduction of constituent quark masses in the presence of quark densities. QCD confinement ensures that net color charge does not appear over large volumes. The general QM grand potential can be parametrized as the following form
~\cite{Alford:2013aca, Zhang:2020jmb}:  
\begin{equation}\begin{aligned}
\Omega=&-\frac{\xi_4a_4}{4\pi^2}\mu^4- 
\frac{ a_2}{\pi^2}  
\mu^2+B\,,
\label{omega_mu}
\end{aligned}\end{equation}
where $\mu$ denotes the chemical potential of QM. The first term describes a free quark gas with additional QCD corrections, with $a_4=1$ corresponding to no corrections and $a_4\ll 1$ corresponding to large corrections~\cite{Fraga:2001id,Alford:2004pf,Weissenborn:2011qu}. The value of $\xi_4$ depends on the flavor composition of QM. For  $ud$QM, $\xi_4= [ (1/3)^{4/3}+ (2/3)^{4/3}]^{-3}\approx1.86$, while for SQM $\xi_4=3$. The $\mu^2$ term arises from  the quark bare mass or possible color superconductivity effect induced by the residual attractive interaction. In this work, we neglect color superconductivity for simplicity, and thus $a_2=0$ for $ud$QM due to the negligible small mass of $u$, $d$ quarks.
$B$ is the effective bag constant that accounts for the nonperturbative feedback of dense quark gases on the QCD vacuum. As the flavor symmetry  is badly broken, $B$ can have  different values for $ud$QM and SQM~\cite{Holdom:2017gdc}.

For later convenience, we define $\lambda=a_2/\sqrt{\xi_4 a_4}$, so that $\lambda=0$ for $ud$QM and $\lambda=-\sqrt{3}m_s^2/(4\sqrt{ a_4})$ for SQM~\cite{Zhang:2020jmb}.  The finite strange quark mass $m_s=95\pm 5 \rm \, MeV$~\cite{ParticleDataGroup:2014cgo}, and we choose $m_s=95 \rm \, MeV$ as its benchmark value.
The general QM EOS is then derived as~\cite{Zhang:2020jmb, Zhang:2021fla, Zhang:2021iah} 
\be
P=\frac{1}{3}(\rho-4B)+ \frac{4\lambda^2}{9\pi^2}\left(-1+\text{sgn}(\lambda) \sqrt{1+3\pi^2 \frac{(\rho-B)}{\lambda^2}}\right).
\label{eos_tot}
\ee 
The EOS of $ud$QM reduces to $P=\frac{1}{3}(\rho-4B)$. The EOS for SQM receives additional corrections from finite $\lambda$, and is softer than the $ud$QM EOS of the same bag constant. 
The energy per baryon number and chemical potential for the interacting QM are~\cite{Zhang:2020jmb}
\begin{eqnarray}
\varepsilon_{Q}&=&\frac{3\sqrt{2} \pi}{(\xi_4 a_4)^{1/4}}\frac{ {B}^{1/4}}{\sqrt{(\lambda^2/B+\pi^2)^{1/2}+\lambda/ \sqrt{B}}}\,, \label{EA} \\
\mu_{\rm Q}&=&\frac{3\sqrt{2}}{(a_4 \xi_4)^{1/4}}\sqrt{[(P+B)\pi^2+\lambda^2]^{1/2}-\lambda}\,.\label{muP_analy}
\end{eqnarray}
The properties of QM are fully determined by the two parameters $(B, a_4)$.

The parameter space of $(B,a_4)$ are constrained by the absolute stability condition of QM at zero pressure. 
Since the most stable nucleus, ${}^{56}\text{Fe}$, has an energy per baryon number of $\varepsilon_{\rm Fe}\approx 930\,$MeV, the stability condition $\varepsilon_{\text{Q}}\leq \varepsilon_{\rm Fe}$ imposes a lower bound on $a_4$,
with 
$a_{4,\rm  min}^{ud\text{QM}}\approx 174\pi^2 B/\varepsilon_{\rm Fe}^4$ 
and  $a_{4,\rm  min}^{\text{SQM}}=3(36\pi^2 B+3 m_s^2 \varepsilon_{\rm Fe}^2)/{\varepsilon_{\rm Fe}^4}$.  In the context of $ud$QM hypothesis, to not ruin the stability of ordinary nuclei, it is safe to have the minimum baryon number of $ud$QM larger than 300, corresponding to $\varepsilon_Q\gtrsim \varepsilon_{900}\equiv 900 \rm\, MeV$~\cite{Holdom:2017gdc}. This sets an upper bound for $a_4$: $a_{4, \rm max}^{ud\text{QM}}=a_{4, \rm min}^{ud\text{QM}} ( \varepsilon_{\rm Fe}\to  \varepsilon_{900}).$   Requiring $a_{4,\rm min}^{ud\text{QM}}\leq 1$ set an upper bound for $ud$QM bag constant $B\leq 56.8\rm\, MeV/fm^3$. 
In the context of the SQM hypothesis, $ud$QM is unstable by definition, and this sets an upper bound of $a_{4,\rm  max}^{\text{SQM}} =a_{4,\rm  min}^{ud\text{QM}}$, under the conventional assumption of a flavour-independent bag constant. Requiring $a_{4, \rm max}^{\text{SQM}} \gtrsim a_{4,\rm min}^{\text{SQM}}$ and $a_{4, \rm min}^{\text{SQM}}\leq 1$ together constrain $B$ to a window $B \in [15.5, 82.2]\rm\, MeV/fm^3$.
Thus, we choose a benchmark set $B=(20, 35, 50) \rm \, MeV/fm^3\approx ( 111^4, 128^4, 140^4)  \rm \, MeV^4$ for both $ud$QM and SQM. 
The allowed range of $a_4$ is given by $a_{4,\rm  min}^{ud\text{QM}}\approx(0.35, 0.62, 0.88)$ and $a_{4,\rm  max}^{ud\text{QM}}\approx (0.40, 0.70, 1.0)$ for $ud$QM; $a_{4,\rm  min}^{\text{SQM}} \approx(0.32, 0.49, 0.64)$ and $a_{4,\rm  max}^{\text{SQM}} \approx(0.35, 0.62, 0.88)$ for SQM.

To address the uncertainties associated with the HM EOS, we perform the analysis for eleven  representative models, including APR, BL, GM1, SLy4, SLy9, DDH$\delta$, Sk13, Sk14, Sk15, SKa, and SKb~\cite{Typel:2013rza, Potekhin:2013qqa,footnotePREX}. The corresponding EoSs and chemical potentials data are taken from Ref.~\cite{Compose}.
We find that CrSs with these hadronic models are not significantly different from each other in terms of their stellar properties such as masses and radii (See Appendix A); thus we take APR as the representative benchmark in this paper.


 \begin{figure}[htb]
\centering
\includegraphics[width=8cm]{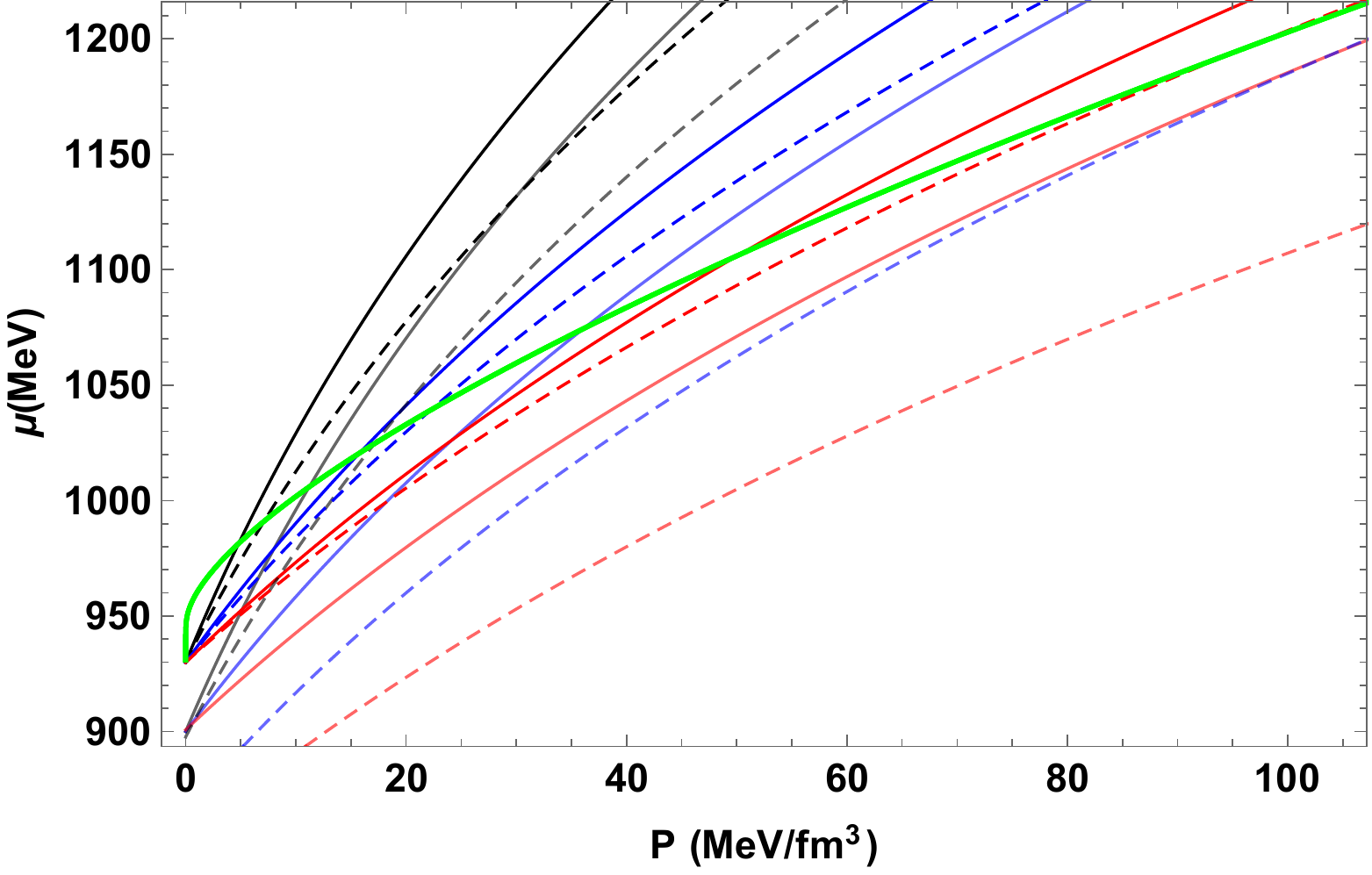}   
\includegraphics[width=8cm]{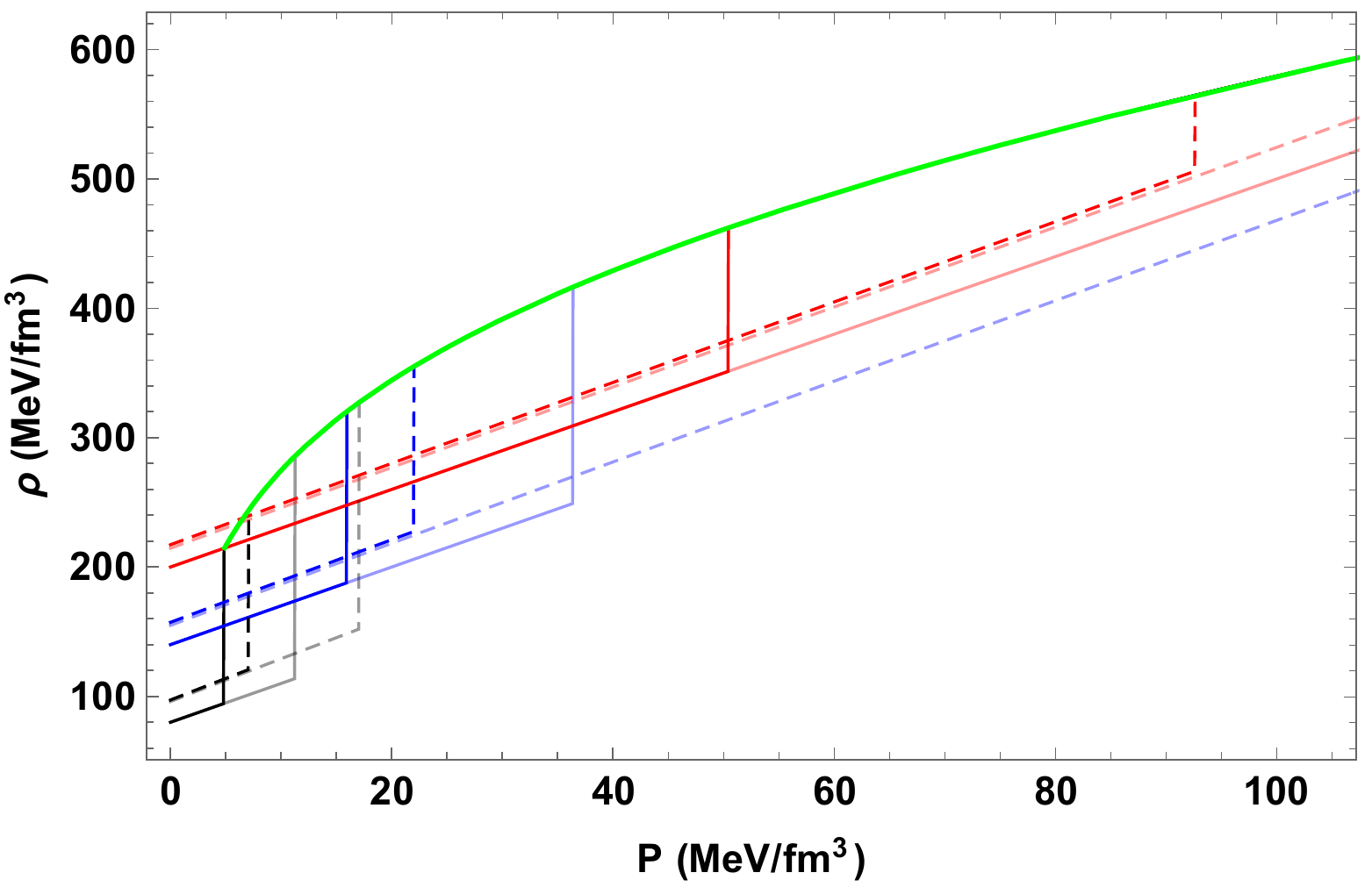}   
 \caption{(top) $\mu (P)$ of the two matter phases and (bottom) $\rho (P)$ (bottom) of cross stars. APR EOS (green) is the representative benchmark for HM. $ud$QM (solid) and SQM (dashed) sample $B=20$ (black), 35\,(blue), 50\,(red)\,$\rm MeV/fm^3$, and $a_4=a_{4,\rm min}$, $a_{4,\rm max}$ from the darker to lighter color. 
 }  
 \label{fig_rhomu}
\end{figure} 

The top panel of Fig.~\ref{fig_rhomu} displays $\mu(P)$ for the two types of QM with the aforementioned benchmark sets of the bag constant $B$, and $\mu(P)$  for APR hadronic EOS~\cite{Compose}.  At low pressure,  $\mu_Q$ is smaller than the hadronic chemical potential $\mu_H$ under the QM hypothesis. As $P$ increases, both chemical potentials go larger. If $\mu_Q$ increases faster with $P$ than $\mu_H$, the two $\mu(P)$ curve may cross at some critical pressure $P_{\rm cr}$. For both cases, $\mu_Q$ increases faster for a smaller $a_4$ or $B$ as expected from Eq.~(\ref{muP_analy}), and so the crossing takes place at a smaller $P_{\rm cr}$.
For SQM, a softer EOS results in a more slowly increasing $\mu_Q$ than $ud$QM, and then a larger value of $P_{\rm cr}$ in general. Thus, to realize CrSs within the stable branch below the maximum mass, the $ud$QM hypothesis allows a larger parameter space.

We depict the combined EOSs in the bottom row of Fig.~\ref{fig_rhomu}. 
At low pressure, the EOS is given by Eq.~(\ref{eos_tot}) for the more stable QM phase. Once the crossing of chemical potentials takes place within the interested range, the energy density jumps from $\rho_Q$ to $\rho_H$ by a finite amount $\Delta \rho$ at $P_{\rm cr}$, indicating a first-order phase transition. 
At higher pressure, HM becomes more stable and the EOSs all merge into the benchmark model.
The possibility of a first-order phase transition has also been considered for conventional hybrid star, while  for CrSs the order of phases are naturally inverted. 
It turns out that there is a large overlap of the parameters they considered and the CrS model we explore here, so the effects of the inverted structure can be more clearly seen.
These cover a large empirical parameter space and are consistent with recent studies on the sound speed~\cite{Landry:2020vaw, Legred:2021hdx, Altiparmak:2022bke,Ecker:2022xxj} (see Appendix B).

\section{Astrophysical Implications}
To obtain the configuration of CrSs, we incorporate the combined EOS of the two matter phases into the Tolman-Oppenheimer-Volkov (TOV) equation~\cite{Oppenheimer:1939ne,Tolman:1939jz}
 \bea
 \begin{aligned}
{dP(r)\over dr}&=-{\left[m(r)+4\pi r^3P(r)\right]\left[\rho(r)+P(r)\right]\over r(r-2m(r))}\,,\,\,\\
{dm(r)\over dr}&=4\pi\rho(r)r^2\,,
\end{aligned}
\label{eq:tov}
\eea
where the profiles $P(r)$ and $m(r)$ are solved as functions of the center pressure $P_{\rm center}$. The radius $R$ and physical mass $M$ of the compact stars are determined by $P(R)=0$ and $M=m(R)$, respectively. 
To compare with gravitational wave observations, we can further compute the dimensionless tidal deformability $\Lambda=2k_2/(3C^5)$, where  $C=M/R$ is the compactness and $k_2$ is the Love number that characterizes the stars' response to external disturbances~\cite{AELove,Hinderer:2007mb,Hinderer:2009ca,Postnikov:2010yn}. 
The Love number $k_2$ can be determined by solving a function $y(r)$ from a specific differential equation \cite{Postnikov:2010yn} and the TOV equation Eq.~(\ref{eq:tov}), with the boundary condition $y(0)=2$. For CrSs, the matching condition~\cite{Damour:2009vw,Takatsy:2020bnx} $y(r_{d}^+) - y(r_{d}^-) = -4\pi r_{d}^3 \Delta \rho_d /(m(r_{d})+4\pi r_{d}^3 P(r_{d}))$ should be imposed at $r_d$ (i.e., the hadronic core radius $r_{\rm cr}$ and the CrS radius $R$), where an energy density jump $\Delta \rho_d$ occurs.

\begin{figure}[h]
\centering
\includegraphics[width=8.8cm]{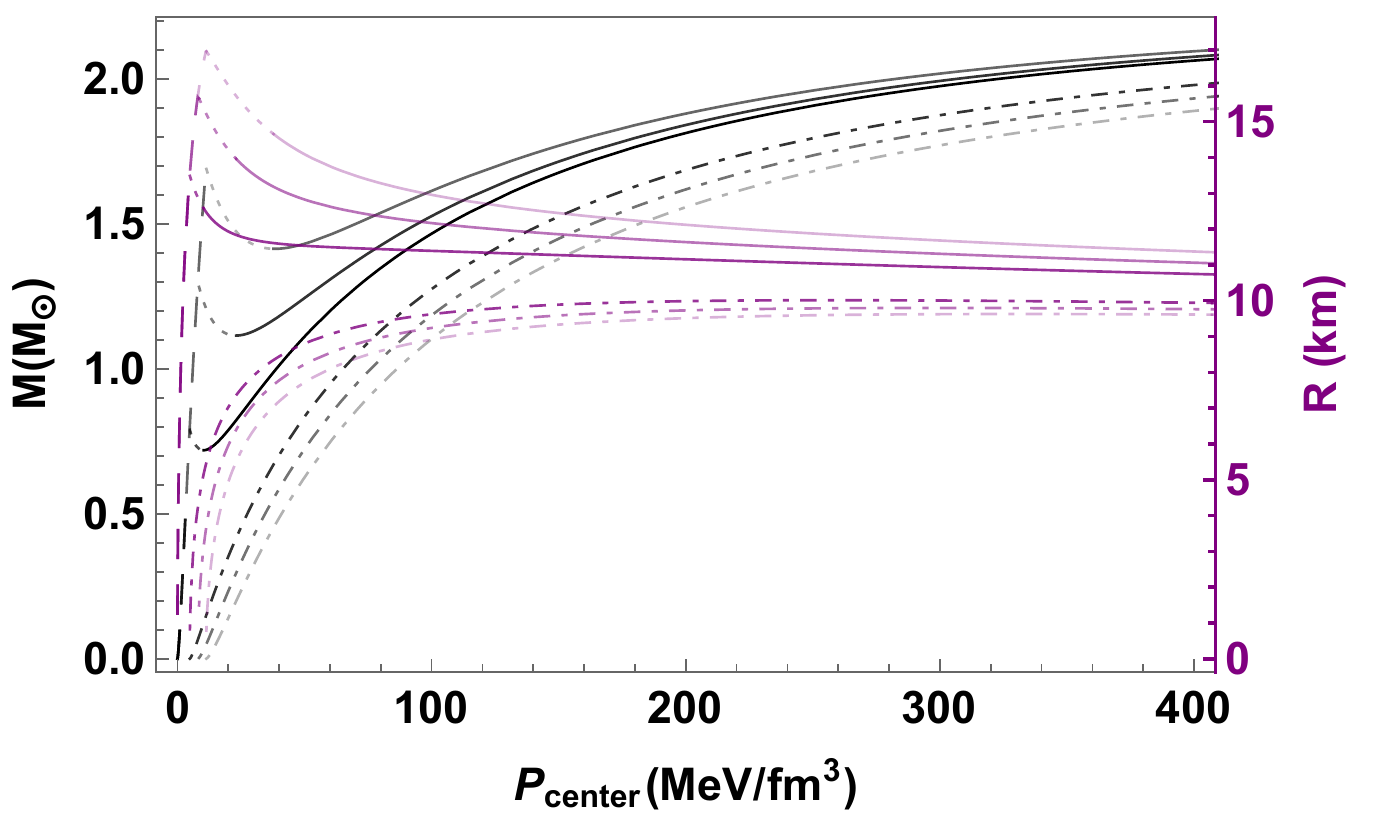}   
 \caption{Mass (black, left axis) and radius (purple, right axis) versus center pressure $P_{\rm center}$ for CrSs with APR	 and $ud$QM of $B=20 \,\rm MeV/fm^3$,  $a_4\approx 0.35, 0.38, 0.40$ from the darker to lighter color. 
Dashed lines denote pure $ud$QSs.  Solid lines denote the stable CrSs with dotted segments denoting the unstable ones. The dot-dashed lines denote the hadronic cores with the radius $r_{\rm cr}$ and mass $m(r_{\rm cr})$.}
 \label{fig_MRP}
\end{figure}

To elaborate the explicit stellar structure of CrSs, we display the masses and radii of the CrSs and their hadronic cores as functions of the centre pressure in Fig.~\ref{fig_MRP}. Here, we choose a benchmark model with the hadronic core described by the APR EOS and the $ud$QM crust with $B=20\rm\, MeV/fm^3$ and different values of $a_4$. CrSs with a SQM crust share similar behaviour.  As a general feature, the compact stars are pure QSs at low $P_{\rm center}$, and transit to CrSs at high $P_{\rm center}$. Right above the transition pressure $P_{\rm cr}$,  there may exist an unstable region, where $M$ decreases as $P$ increases. When $M$ approaches the maximum mass, the HM core grows large and the QM crust  takes only a small fraction with $\sim 0.1\, M_\odot$ in mass and $\sim 1\,$km in width.

Regarding the dependence on QM properties, we see that, for a given $B$, a lower $P_{\rm cr}$ results in a larger mass and radius for the hadronic core and the opposite effect on the overall mass and radius, thus altogether map to a larger mass and radius fraction for the hadronic core. Also, the unstable region grows larger for increasing $a_4$, which maps to the large $\Delta \rho/\rho(P_{\rm cr})$ regime referring to Fig.~\ref{fig_rhomu}. Such a map is a general feature for compacts stars involving hybrid structures~\cite{Lindblom:1998dp, Alford:2013aca}.

\begin{figure*}[htb ]
\centering
\includegraphics[width=8cm]{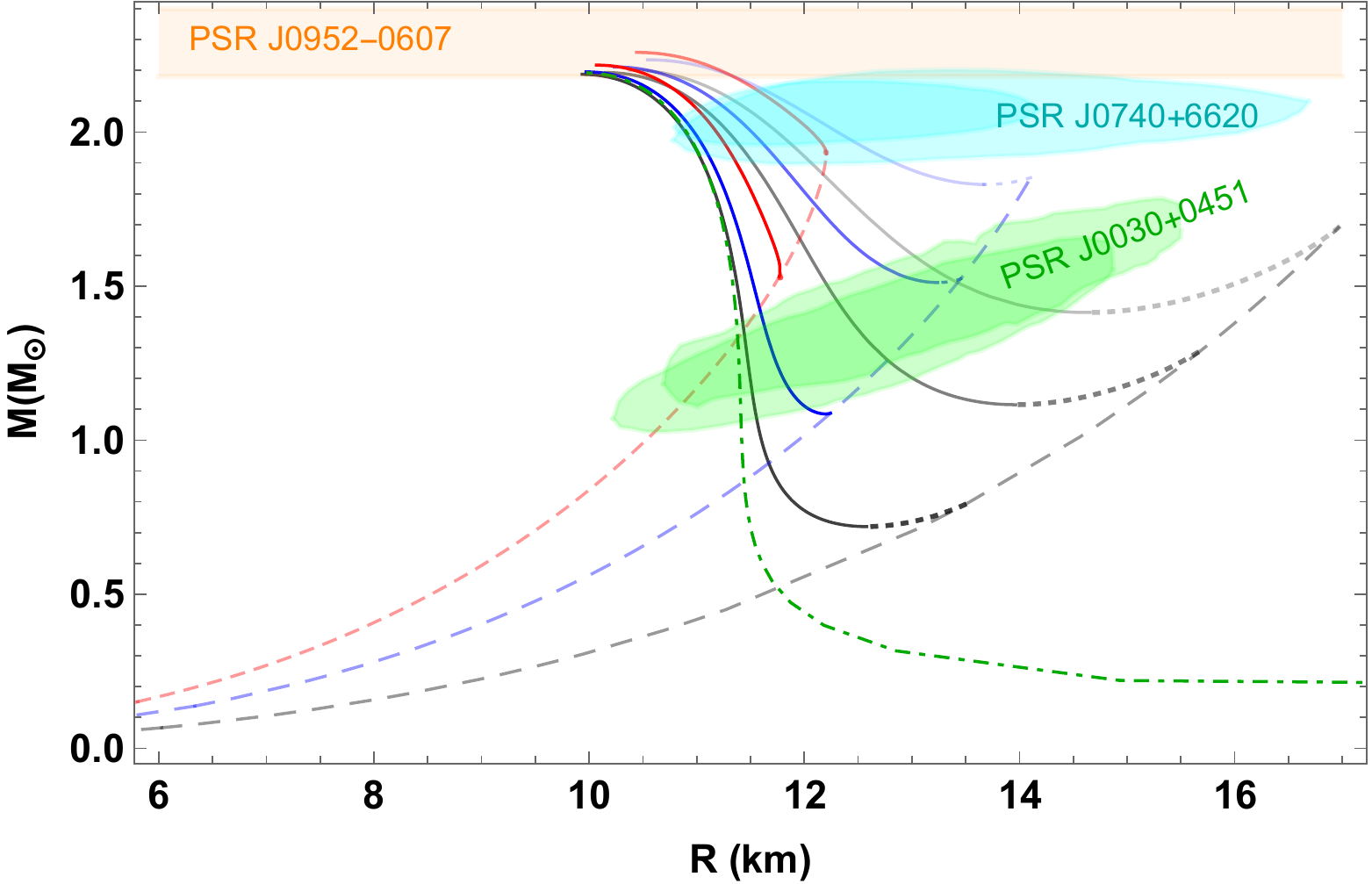}   
\includegraphics[width=8cm]{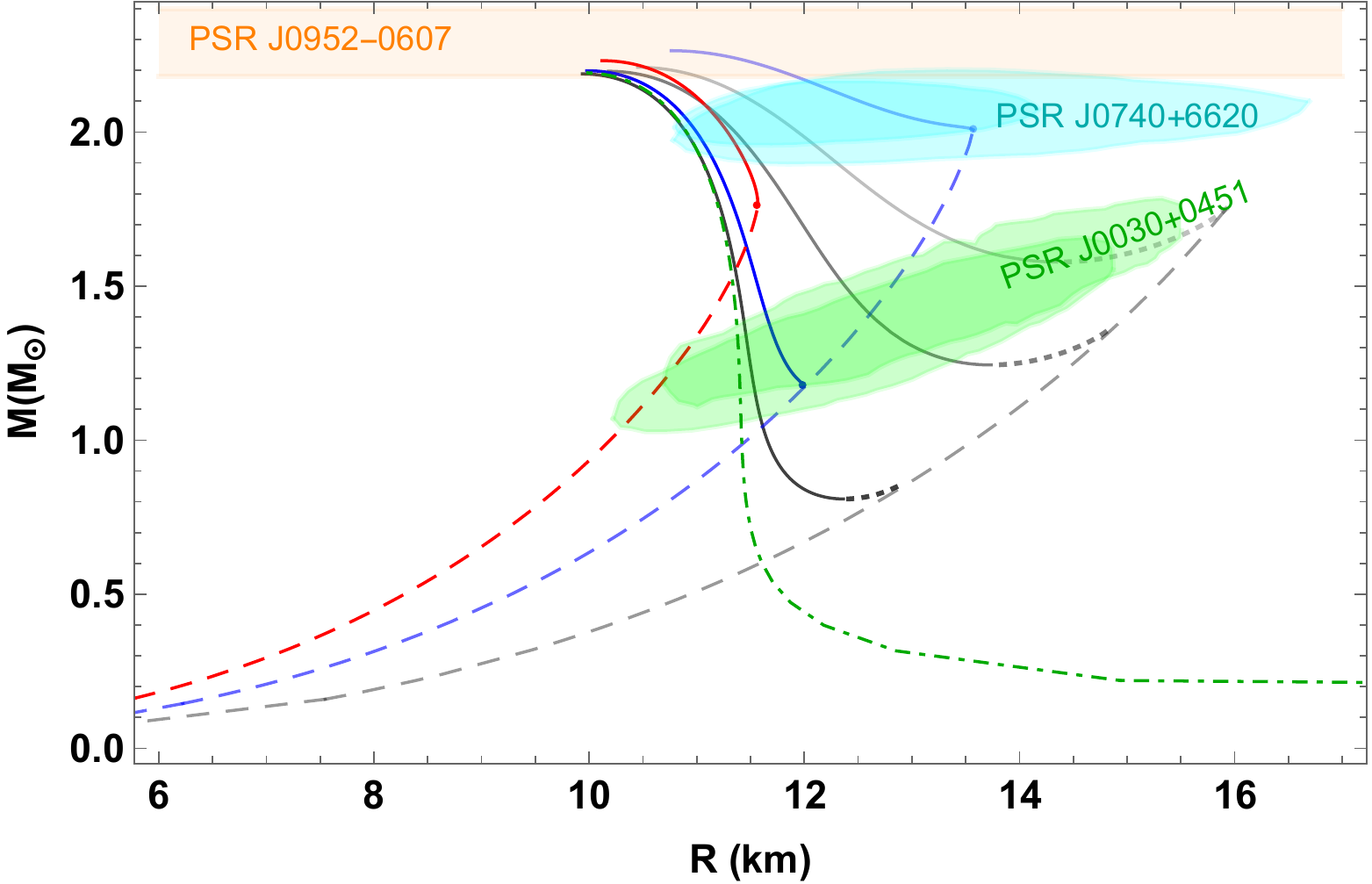}   
\includegraphics[width=8cm]{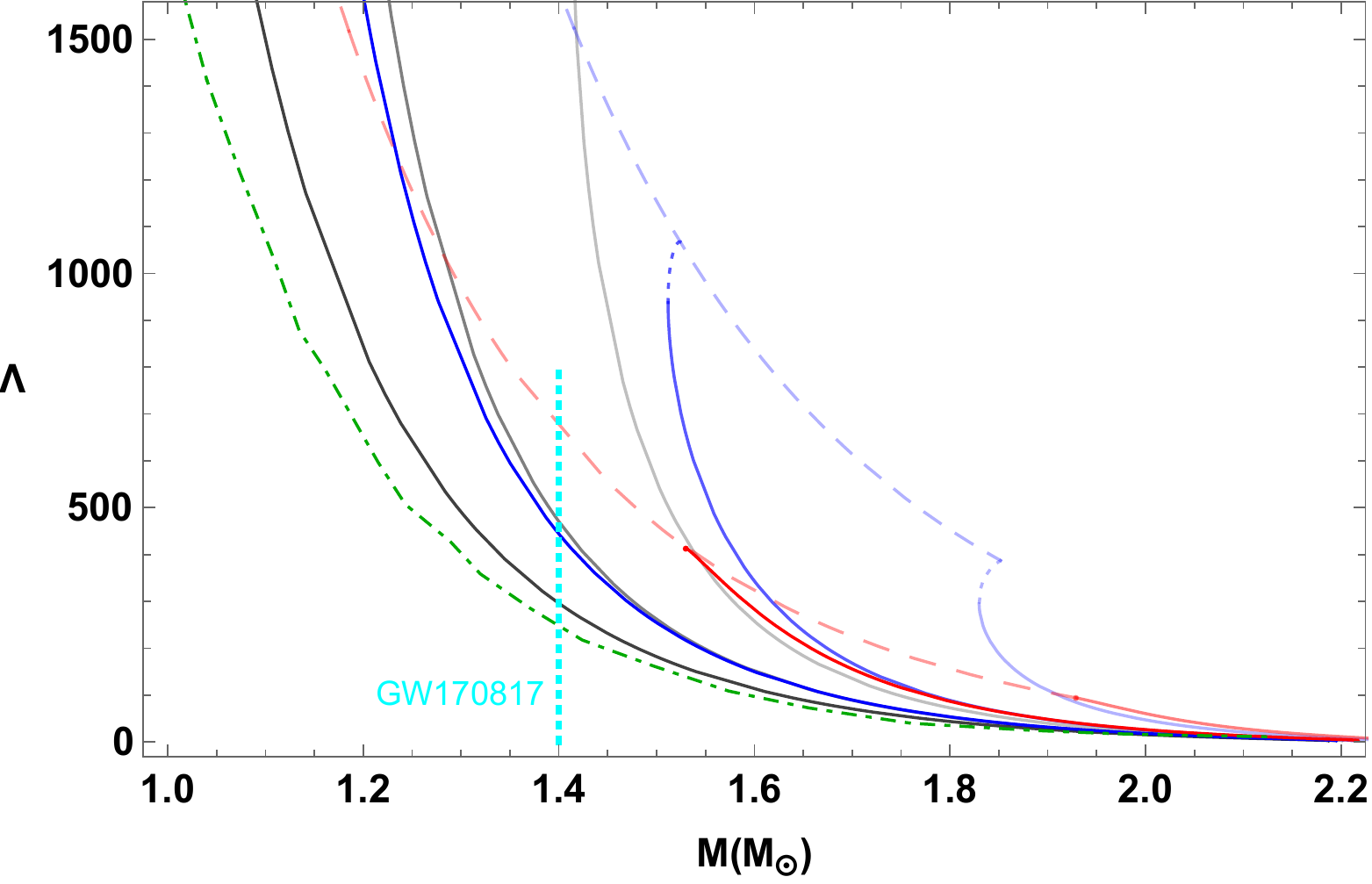} 
\includegraphics[width=8cm]{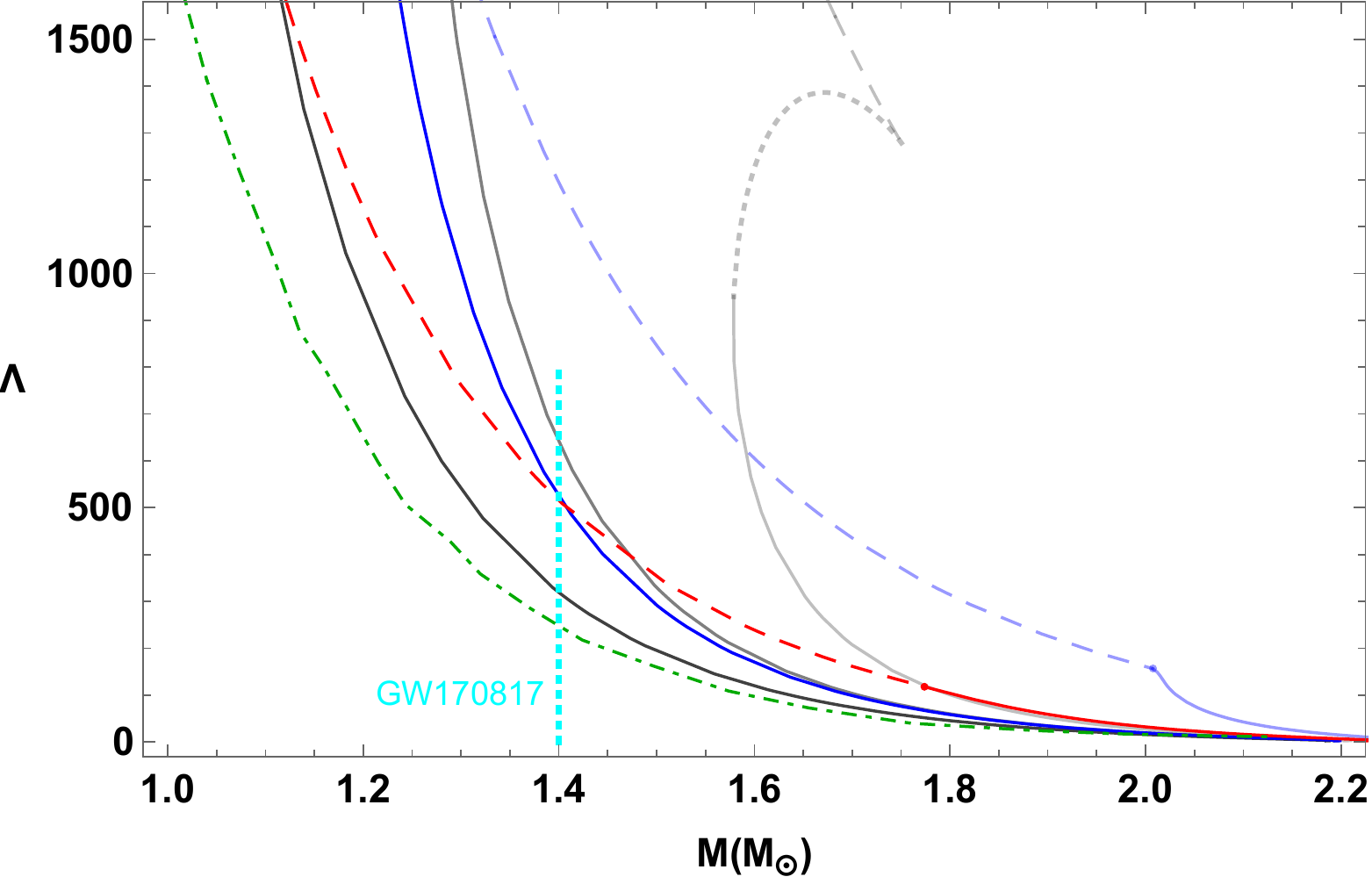}  
 \caption{M-R (top) and $\Lambda$-$M$ (bottom) of cross stars (solid lines) with $ud$QM (left) and SQM (right), sampling $B=20$ (black), 35\,(blue), 50\,(red)\,$\rm MeV/fm^3$, and $a_4=a_{4,\rm min}$, $(a_{4,\rm min}+a_{4,\rm max})/2$, $a_{4,\rm max}$ from the darker to lighter color. All cases are truncated to the maximum mass points, and the cases with no CrS branch are dropped.  APR EOS is used for the hadronic composition, with the green dot-dashed curve denoting pure hadronic star with APR EOS. Dashed lines denote pure QSs, while solid and dotted lines denote stable and unstable CrSs. Shaded regions are constraints with $90\%$ credibility from the NICER mission PSR J0030+0451 (green colored) \cite{NICER1,NICER2}, PSR J0740+6620 (cyan colored)\cite{NICER3,NICER4}, and PSR J0952-0607~\cite{Romani:2022jhd} (orange colored). The cyan-dotted vertical line denote GW170817's $\Lambda(1.4 M_{\odot})\lesssim 800$ constraint~\cite{Abbott:2018wiz}. 
}
 \label{fig_all}
\end{figure*}

We depict the mass-radius relation and the corresponding tidal deformabilities of CrSs with a more general benchmark set in Fig.~\ref{fig_all}. We see that, in general, the masses of CrSs are always larger than the corresponding HSs of the same radii, while the radii and tidal deformabilities of CrSs are in between those for the corresponding QSs and HSs of the same $M$.  As $M$ increases, their radii and tidal deformabilities decrease and approach those of the pure HS due to the growth of the hadronic core. This helps to satisfy the GW170817 constraint
in comparison to pure QSs of the same $B$. On the other hand, the maximum mass ($M_{\rm TOV}$) and radii of CrSs are lifted by the quark matter crust compared to pure HSs, which helps meet the NICER J0740+6620 constraint 
and observations of heavy pulsars.  The advantage of CrSs is then clear: the interplay between the HM and QM compositions helps to reconcile astrophysical constraints at low and high masses, opening vastly more space for soft hadronic EOSs or stiff quark matter EOSs.

More explicitly, the depicted maximum masses of CrSs all meet the two solar mass constraints from pulsar mass measurements, in particular,  $2.01^{+0.17}_{-0.11} M_{\odot}$ for J0348+0432~\cite{Antoniadis:2013pzd}, $2.14^{+0.20}_{-0.18} M_{\odot}$  for J0740+6620~\cite{NANOGrav:2019jur} and $2.35^{+0.17}_{-0.17} M_{\odot}$ for PSR J0952-0607~\cite{Romani:2022jhd}.
For the rest, we find two possibilities naturally emerge to reconcile the $M$-$R$ relation constraints of PSR J0740+6620 \cite{NICER3,NICER4}  and  PSR J0030+0451 \cite{NICER1,NICER2} from NICER and the tidal deformability constraint of the GW170817 event. The first one is to have the CrS branch to explain them all. We find that in small $B$  or $a_4$ region with not-large $P_{\rm cr}$, CrSs can account for a sufficiently small tidal deformability at $1.4M_{\odot}$ and a large enough $M_{\rm TOV}$ simultaneously.   
The other possibility is to have lower mass QSs explaining the PSR J0030+0451 and GW170817 constraints and CrSs residing in the large-mass regime meeting the PSR J0740+6620 one. This arises in the large $B$ region with a relatively large $P_{\rm cr}$, as the red lines ($B=50\,\rm MeV/fm^3$) shown in Fig.~\ref{fig_all},
where $M_{\rm TOV}$ of CrS can even exceed that of QS.

For the two types of QM hypothesis, the parameter space for CrSs with a SQM crust is more constrained by the astrophysical observations, especially for hadronic EOSs that are relatively stiffer than APR at low pressure like SLy4 (see Appendix A). This is because the transition usually takes place at a higher $P_{\rm cr}$ for SQM compared to $ud$QM, as shown in Fig.~\ref{fig_rhomu}. Thus, less parameter space exists to realize stable CrSs in the SQM hypothesis in general, and we see fewer solid lines on the right column of Fig~\ref{fig_all} than the left.

We also see ``twin star" configurations, i.e. stars with identical masses but very different radii, in cases of small $B$ and large $a_4$. This is closely related to the existence of the unstable region in Fig.~\ref{fig_MRP} and the large $\Delta{\rho}/\rho(P_{\rm cr})$ for the first-order phase transition as discussed. In contrast to conventional hybrid stars, the larger-radius one in our scenario is a pure QS and the smaller-radius one is a CrS.
For CrSs with $ud$QM (SQM) in the cases shown, the largest radius difference can reach $4.3 \rm \, km$ ($3.1 \rm \, km$)  at $M \sim 1.7 \,\rm M_\odot$ ($1.75 \rm\, M_\odot$).

\section{Conclusions}
To summarize, we have explored a new possibility of a distinct stellar objects, ``Cross stars" (CrSs), with a QM crust and a HM core, based on the hypothesis that QM can be absolutely stable at low pressure. 
We have shown that CrSs can be naturally realized within the stable branch of compact stars and they may play a significant role to reconcile various astrophysical constraints.

It is possible that more crossings exist at higher pressure, yielding a multilayer structure of quark and hadronic matter. The astrophysical implications of such objects deserve further study, including the possibility of explaining the supergiant glitch phenomena, considering the associated first order phase transition decreases the moment of inertia and thus causes a sudden spin-up~\cite{Glendenning:1997fy, Ma:1996vv}.  CrSs can possibly be distinguished from other type of compact stars from future fine-radius measurement~\cite{Li:2022rxw}, or GW asteroseismology~\cite{Passamonti:2005cz,Pratten:2019sed} due to their distinction on nonradial oscillations~\cite{Zhang:2023zth}.
Besides, including color superconductivity effects can increase the stiffness of QM EOSs and then  enlarge the parameter space of CrSs, which may explain the massive secondary component of GW190814~\cite{Abbott:2020khf,Zhang:2020jmb,Kanakis-Pegios:2020kzp,Roupas:2020nua}.

\begin{acknowledgments}
\subsection*{Acknowledgments}
C.~Z. is supported by the Jockey Club Institute for Advanced Study at The Hong Kong University of Science and Technology.  J. R. is supported by the Institute of High Energy Physics, Chinese Academy of Sciences, under Contract No. Y9291220K2. 
\end{acknowledgments}

\setcounter{equation}{0}
\setcounter{figure}{0}
\setcounter{table}{0}
\setcounter{page}{1}
\makeatletter
\renewcommand{\thefigure}{A\arabic{figure}}


\section{Appendix A: Variations of hadronic EOSs}
\label{AppVaryHM}
To demonstrate the robustness of our results against the uncertainties of hadronic EOSs, we consider ten more examples of hadronic EOSs here.  

\begin{figure}[htb]
\centering
\includegraphics[width=8cm]{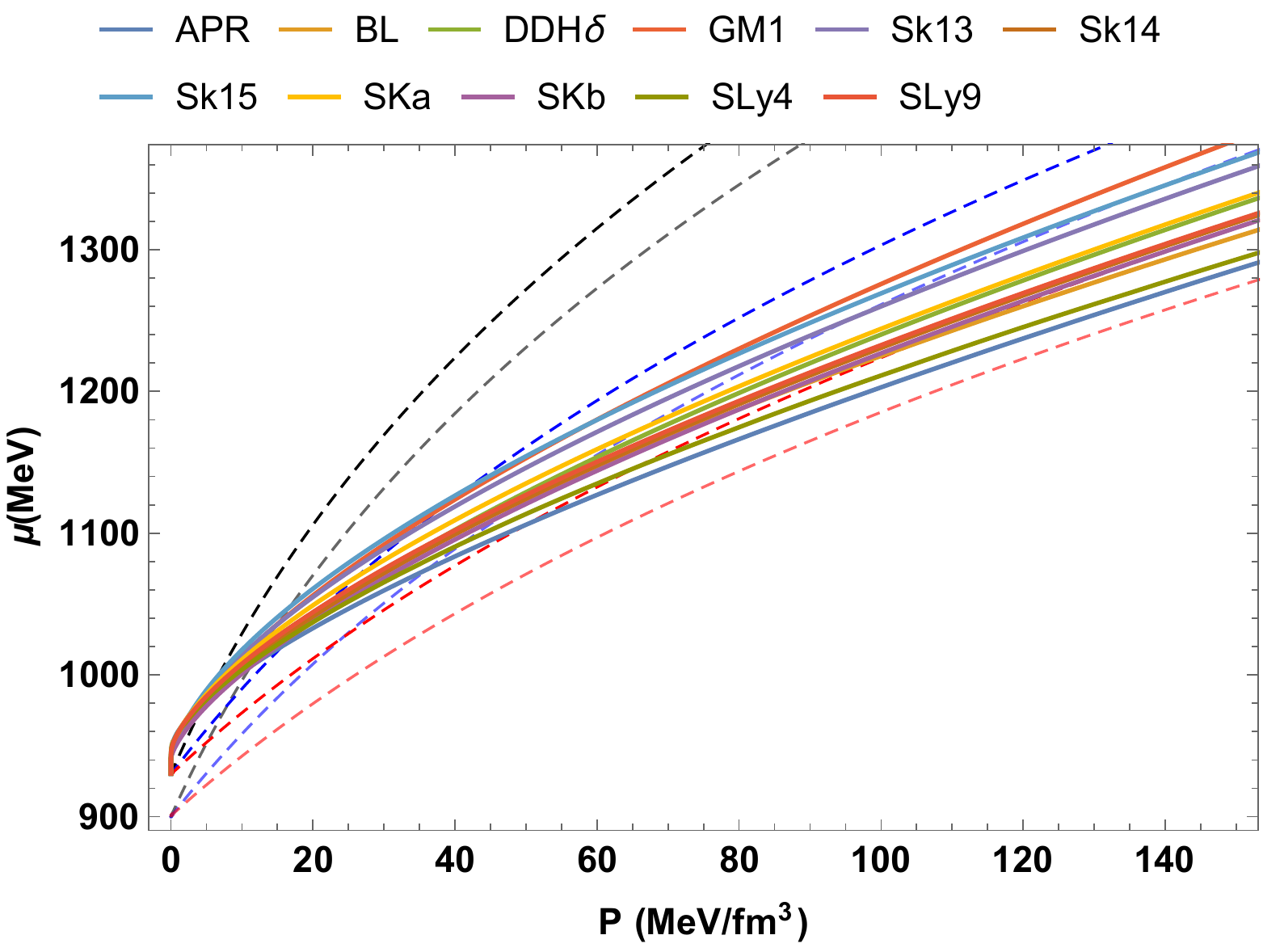}   
\includegraphics[width=8cm]{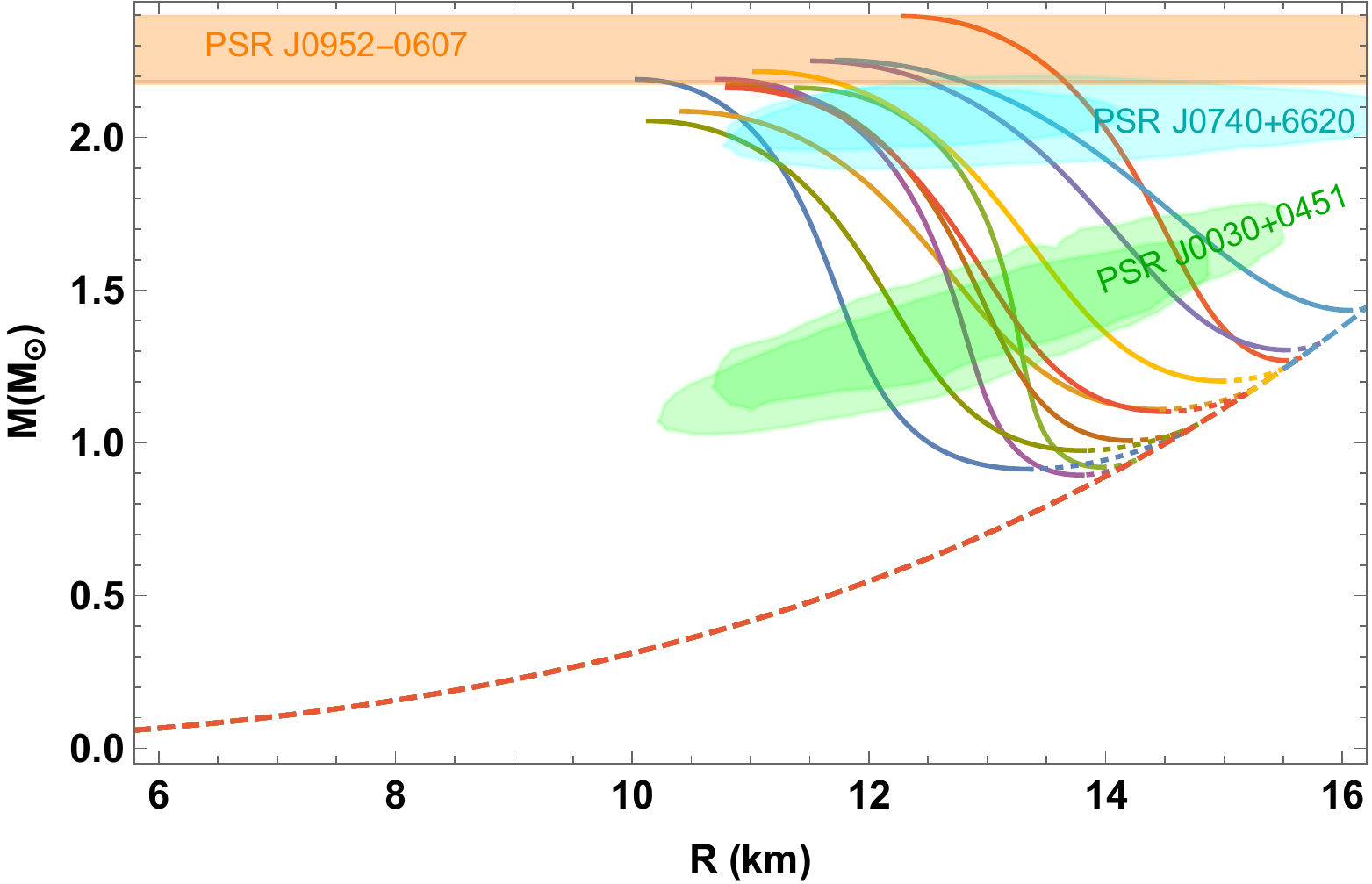}   
\includegraphics[width=8cm]{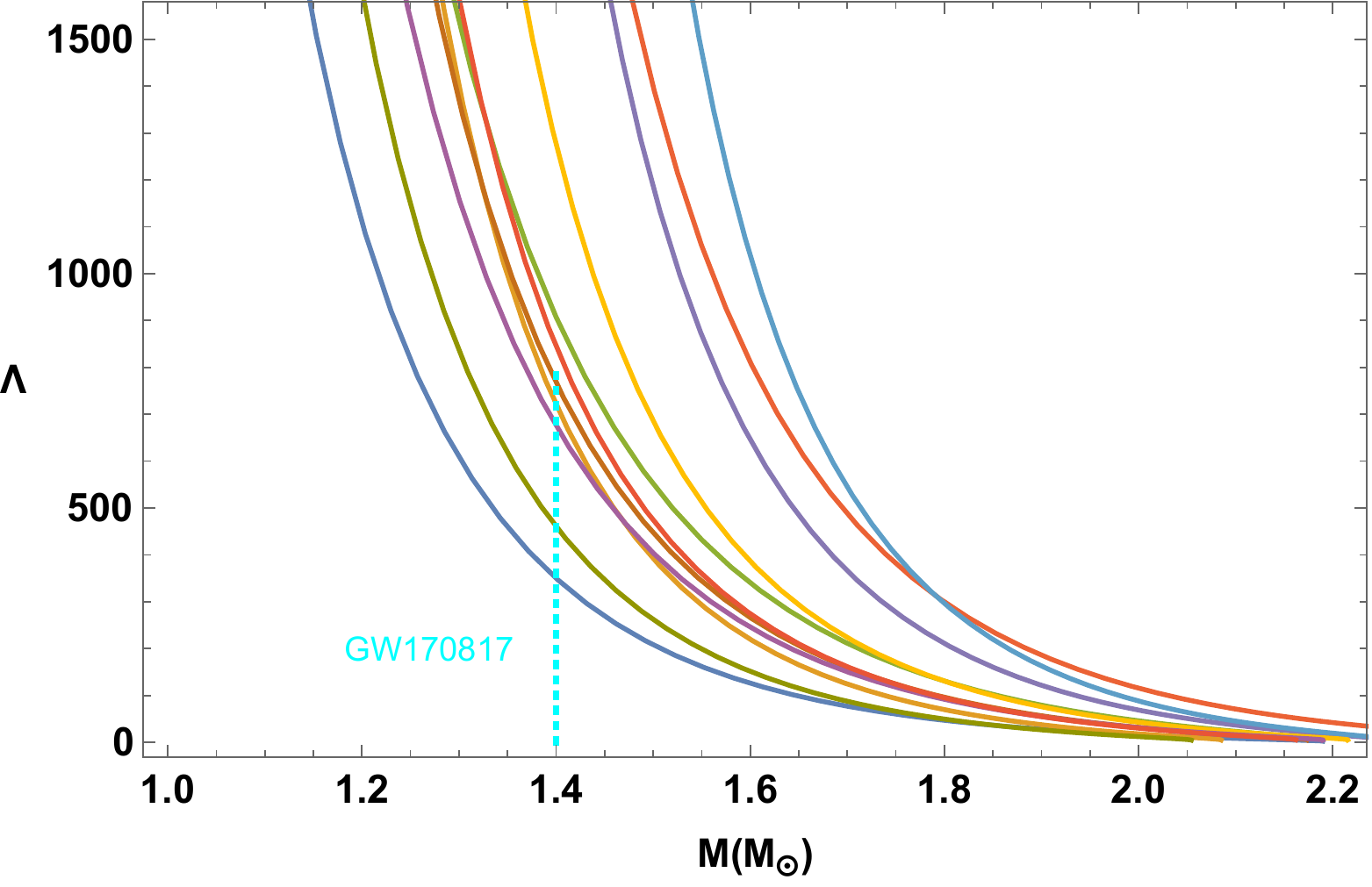}  
 \caption{(Top) $\mu (P)$ of the two matter phases with $ud$QM EOSs same as Fig.~\ref{fig_rhomu} but now with 11 hadronic EOSs.  (Middle) M-R  and (bottom)  $\Lambda$-M of cross stars (solid curves) with $ud$QM of $(B, a_4)=(20\rm \, MeV/fm^3, 0.35)$. The 11 hadronic EOSs are those as the top legends shows. Dashed lines denote pure QSs configurations and the dotted lines denote unstable CrSs configurations. 
The observational constraints follow the same as Fig~\ref{fig_all}.} 
 \label{fig_HS}
\end{figure} 

We display in Fig.~\ref{fig_HS} the masses, radii, and tidal deformabilities of CrSs for the eleven HM EOSs. They have similar $M-R$ and $\Lambda-M$ relations, and all meet the two solar mass constraints. The main distinctions are at $P_{\rm cr}$ and $M_{\rm cr}$. As mentioned in the main text, given the general thermodynamic relation $\mu(P)\approx \mu(0)+\int_0^P dP'\frac{1}{n(P')}\,$, $P_{\rm cr}$ is smaller in the case of a softer HM EOS at low densities. 
For example, the BL EOS is stiffer than SLy4 at both low and high pressure, thus it has a larger $P_{\rm cr}$ and $M_{\rm cr}$. In contrast, APR is softer than SLy4 at low densities but stiffer than SLy4 at high densities. APR then results in a smaller $P_{\rm cr}$ and $M_{\rm cr}$, but with a larger maximum mass $M_{\rm TOV}$ at high densities. Those with stiff hadronic EOS at low densities are also less likely to meet the GW170817 tidal deformability constraints.

We enumerate explicit results of CrSs with SLy4 below in Fig.~\ref{fig_all_SLy}. Compared to CrSs with APR in Fig.~\ref{fig_all}, we see that CrSs with $ud$QM are quite similar. Due to the fact that SLy4 is softer than APR at high pressure, we see CrSs with SLy4 can not easily meet the $M_{\rm TOV}$ constraint of $2.35^{+0.17}_{-0.17} M_{\odot}$ from PSR J0952-0607~\cite{Romani:2022jhd}~\footnote{One can employ CrSs with stiffer HM EOS (like what we did in the main text) or stiffer QM EOS (like QM with color superconductivity) at large pressure to easily satisfy this large $M_{\rm TOV}$ constraint.}, yet it can still easily meet all other constraints such as GW170817 (cyan-colored dashed line), PSR J0740+6620 and PSR J0030+0451 from NICER (shaded regions), plus the $M_{\rm TOV}$ constraints of $2.01^{+0.17}_{-0.11} M_{\odot}$ for J0348+0432 and $2.14^{+0.20}_{-0.18} M_{\odot}$ for J0740+6620, with the help of lift on masses and radii compared to pure hadronic stars with SLy4.

\begin{figure*}[htb!]
\centering
\includegraphics[width=8cm]{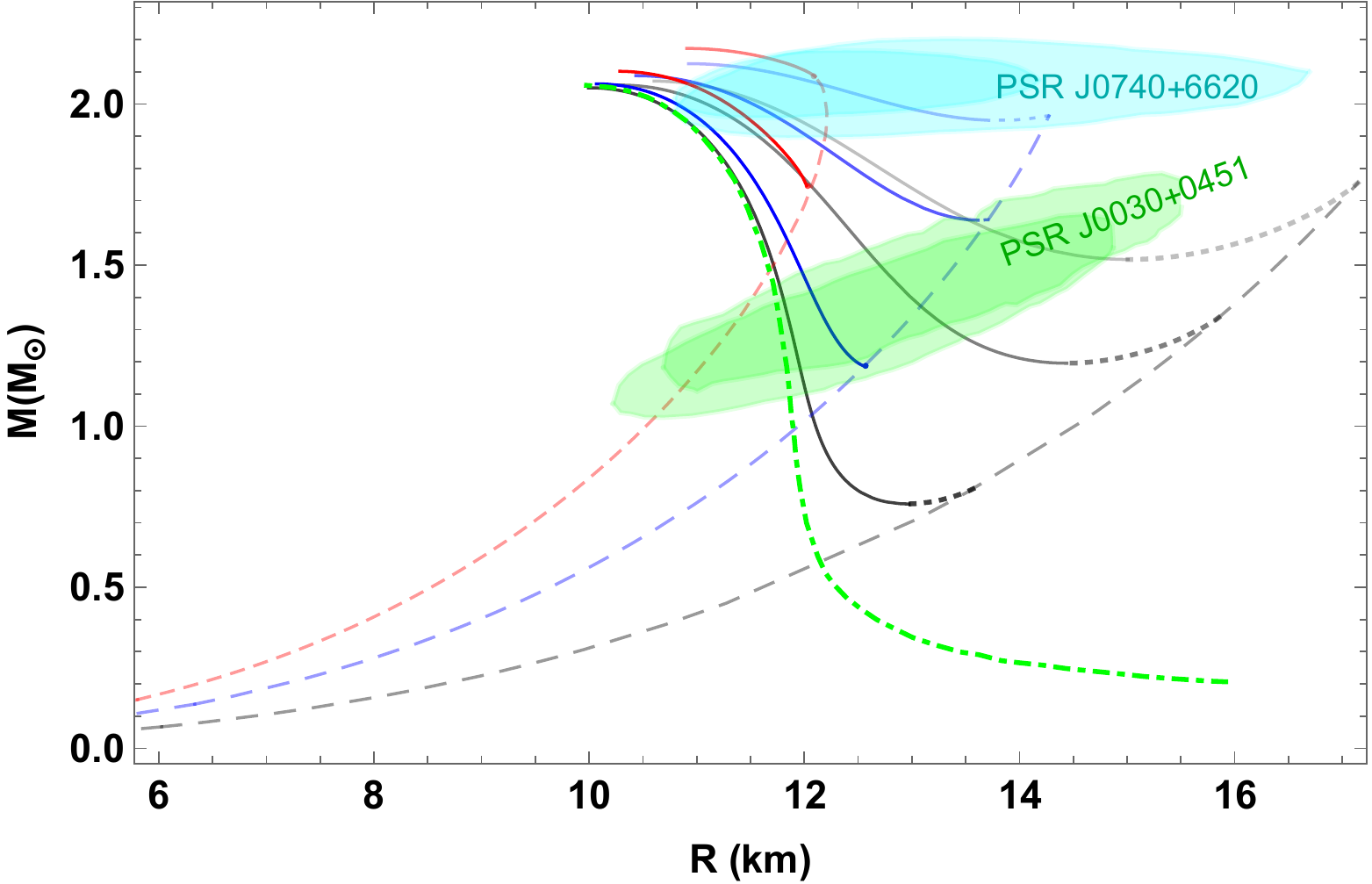}   
\includegraphics[width=8cm]{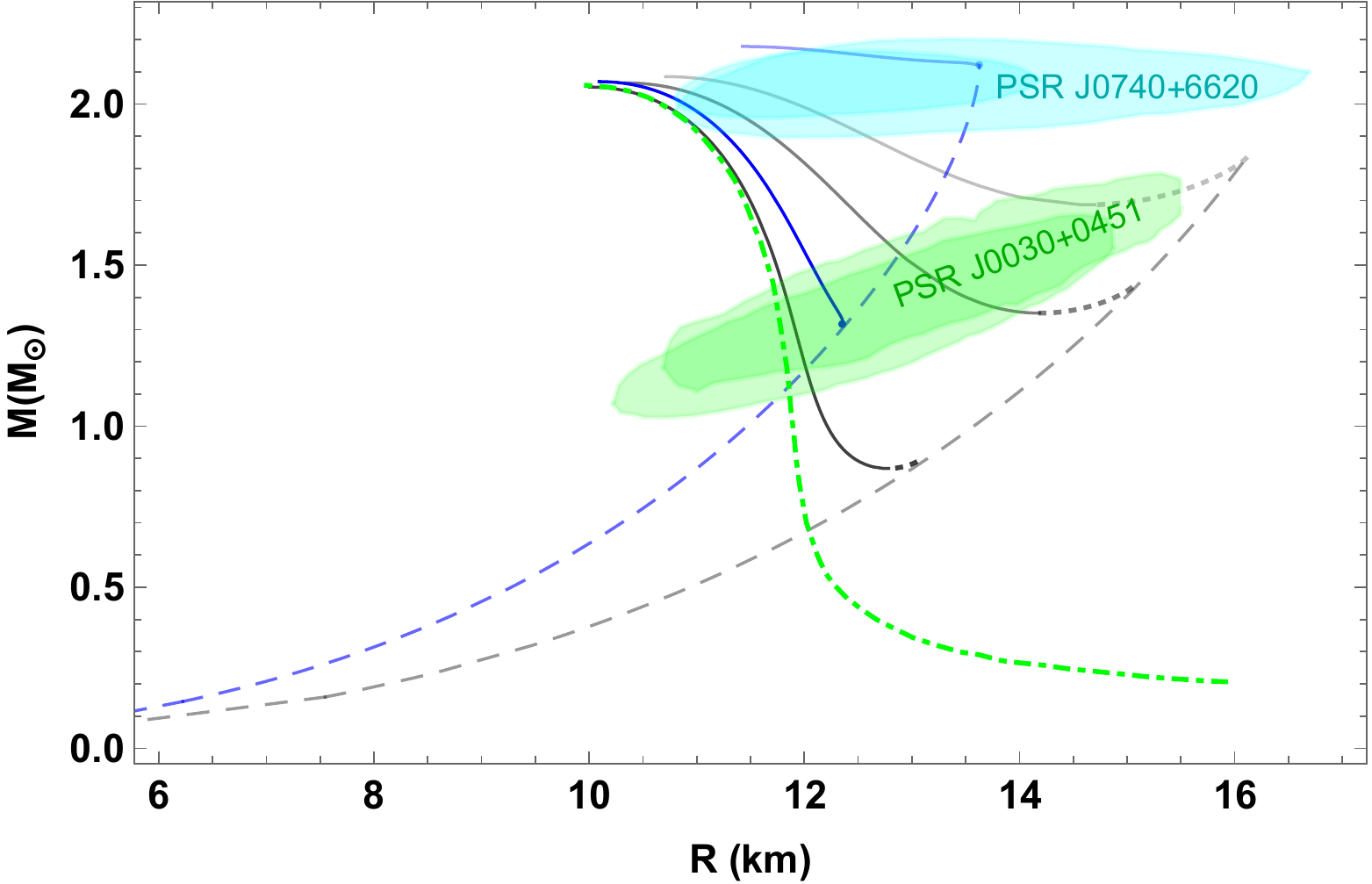}   
\includegraphics[width=8cm]{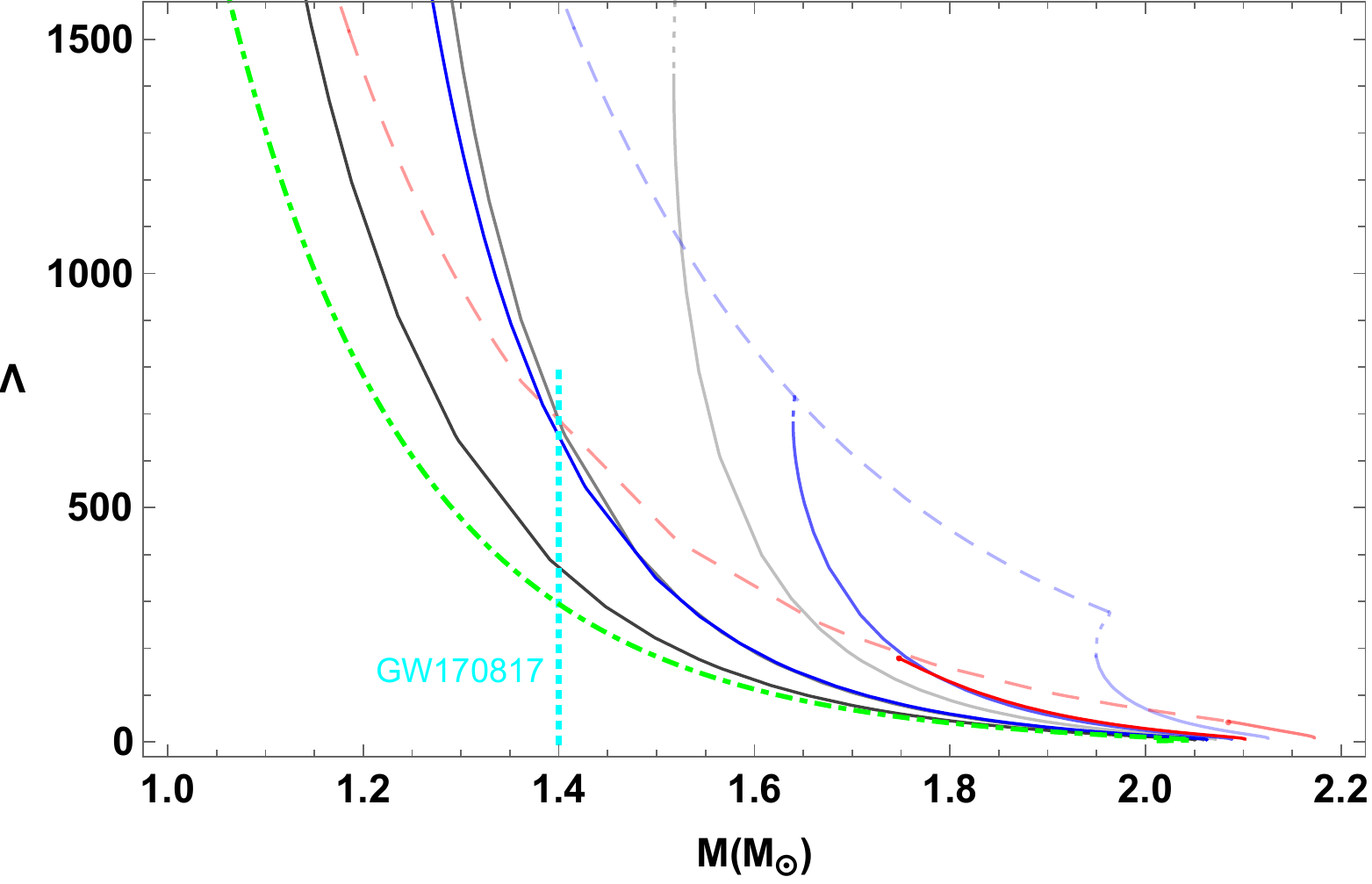} 
\includegraphics[width=8cm]{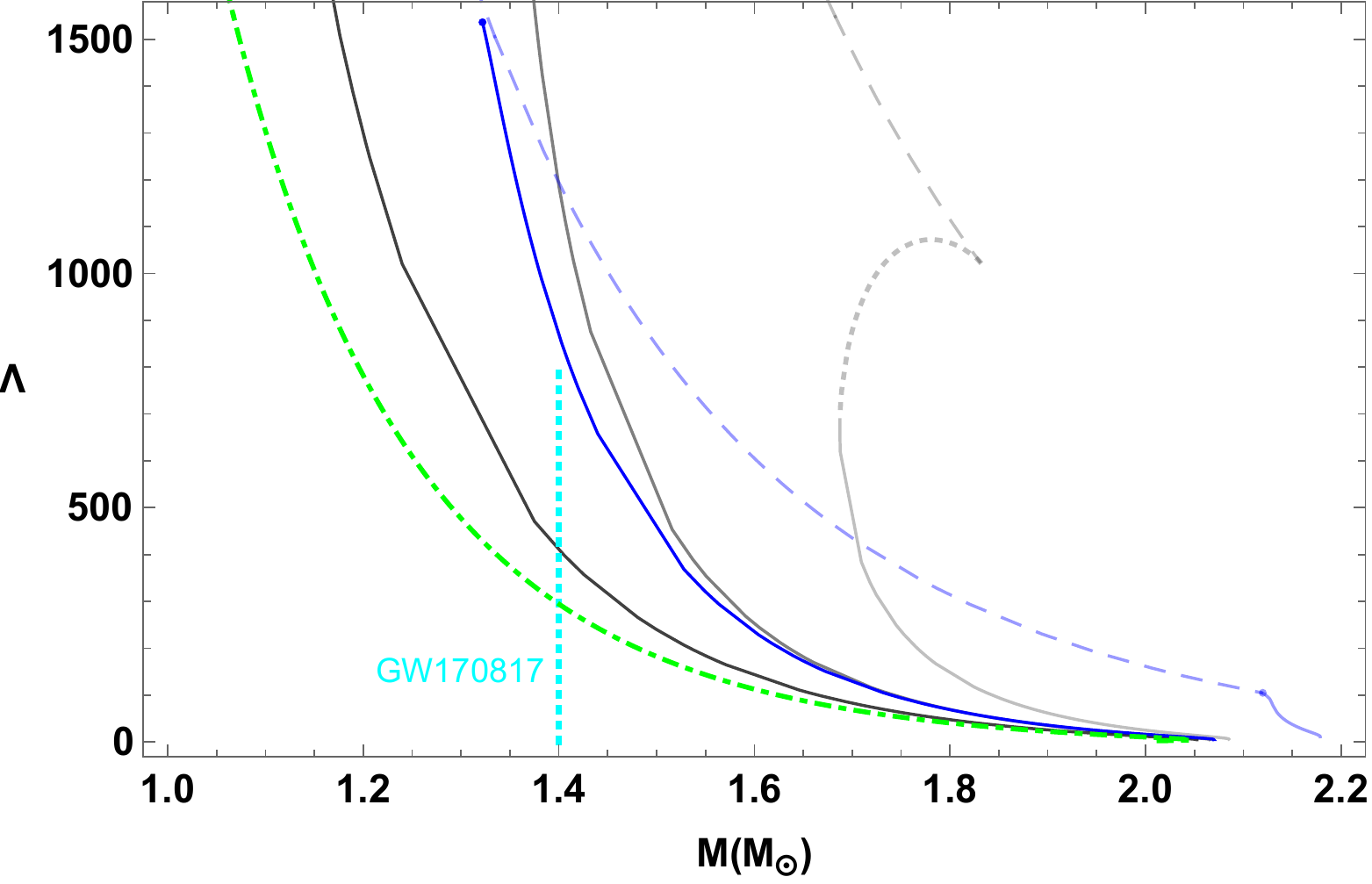}  
 \caption{M-R (top) and $\Lambda$-$M$ (bottom) of Cross stars (solid lines) with $ud$QM (left) and SQM (right). SLy4 EOS is used for hadronic composition, with the green dot-dashed curve denoting pure hadronic star with SLy4 EOS. All other conventions follow the same as Fig~\ref{fig_all}.
}
 \label{fig_all_SLy}
\end{figure*} 

 However, there are noticeable differences for CrSs with SQM, i.e. less parameter space allows the existence of CrSs (no red lines) and to satisfy GW170817 tidal deformability constraints in Fig.~\ref{fig_all_SLy}. These differences are expected considering that SLy4 is stiffer than APR at lower pressure and then results in larger tidal deformabilities and larger $P_{\rm cr}$. The SQM composition has a similar effect so that altogether makes it less likely for CrSs with SQM to exist, and less likely to satisfy GW170817 constraints even if they exist.

\setcounter{equation}{0}
\setcounter{figure}{0}
\setcounter{table}{0}
\setcounter{page}{1}
\makeatletter
\renewcommand{\theequation}{B\arabic{equation}}
\renewcommand{\thefigure}{B\arabic{figure}}

\section{Appendix B: Comparison with conventional hybrid stars}
\label{AppCSS}
The constant-sound-speed (CSS) parametrizations~\cite{Alford:2013aca} are usually used for a model-agnostic description of the EOSs for hybrid stars:
\be
\rho(P) = \left\{\!
\begin{array}{ll}
\rho_{\rm HM}(P) & P<P_{\rm trans} \\
\rho_{\rm trans}+\Delta\rho+c_s^{-2} (P-P_{\rm trans}) & P>P_{\rm trans} 
\end{array}
\right.\ ,
\label{eqn:EoSqm1}
\ee
where $P_{\rm trans}$ is the critical pressure of HM-to-QM phase transition, $\rho_{\rm trans}=\rho_{HM}(P_{\rm trans})$, $\Delta \rho=\rho_{\rm QM}(P_{\rm trans})- \rho_{\rm HM}(P_{\rm trans}) $, and $c_s$ is the sound speed of QM. A first-order phase transition takes place when $\Delta\rho>0$.

The CrSs in contrast have an inverted structure and the EOS has the following form 
 \be
\rho(P) = \left\{\!
\begin{array}{ll}
 \rho_{\rm QM}(P)  & P<P_{\rm trans}  \\
\rho_{\rm HM}(P)    & P \gtrsim P_{\rm trans} 
\end{array}
\right.\ ,
\label{eqn:EoSqm2}
\ee
with  $P_{\rm trans}=P_{\rm cr}$ determined from the crossing of $\mu_Q$ and $\mu_H$. Eq.~(\ref{eqn:EoSqm2}) can be compared with the CSS parameterization in Eq.~(\ref{eqn:EoSqm1}) but with an inverted order of HM and QM. This leads to $\rho_{\rm trans}=\rho_{\rm QM}(P_{\rm trans})$ and $\Delta \rho= \rho_{\rm HM}(P_{\rm trans}) -\rho_{\rm QM}(P_{\rm trans})$.  It turns out that our interacting quark matter model for $ud$QM (SQM) together with APR hadronic EOS in Fig.~\ref{fig_rhomu} cover a large variation of CSS parameters: $P_{\rm trans}=5\sim 96\, (7\sim 93)\,\rm MeV/fm^3$, $\rho_{\rm trans}=94\sim 489\, (121\sim 506)\,\rm MeV/fm^3$, $\Delta \rho=83\sim111\, (59\sim 175 )\,\rm MeV/fm^3$, and correspondingly $\Delta\rho/\rho_{\rm trans}=0.17\sim1.51\, (0.12\sim 1.15)$. 

\begin{figure}[h]
\centering
\includegraphics[width=8 cm]{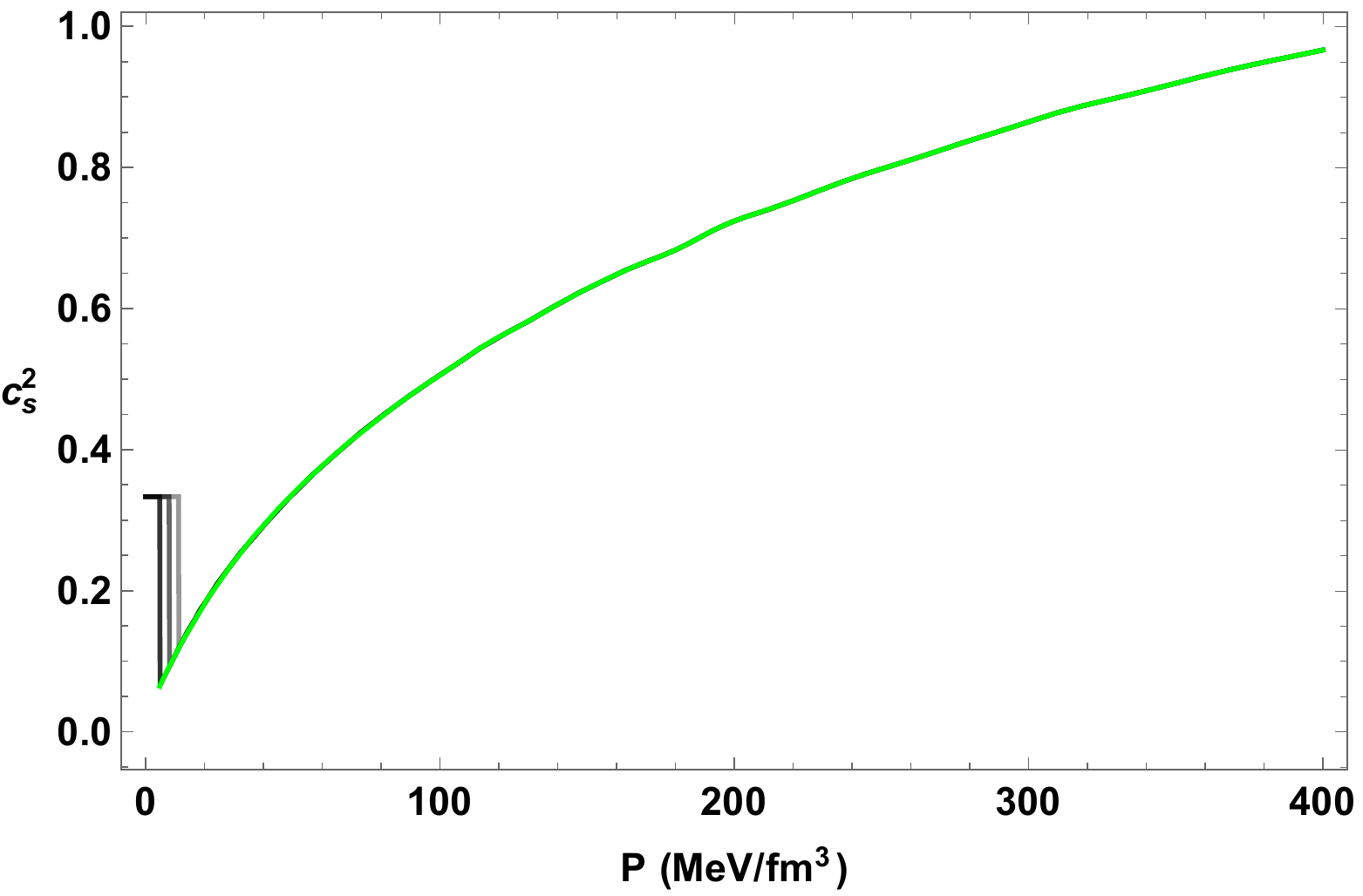}   
 \caption{ $c_s^2$ vs $P$ for CrSs with APR (green) and $ud$QM of $B=20 \,\rm MeV/fm^3$,  $a_4\approx 0.35, 0.38, 0.40$ from the darker to lighter black color. }
 \label{fig_cs2}
\end{figure}

For the sound speed, the QM crust has $c_s^2=1/3$ for $ud$QM and $c_s^2<1/3$ for SQM, while the HM core can have the sound speed violating the conformal speed limit ($c_s^2>1/3$), as shown by Fig.~\ref{fig_cs2} for the same benchmark parameter set in Fig.~\ref{fig_MRP}.

This does not contradict with the recent generic studies of the sound speed in neutron stars~\cite{Legred:2021hdx,Ecker:2022xxj, Altiparmak:2022bke}, which have assumed the low-densities EOS to be hadronic and then do not account for the possibility of absolute stable QM.

\clearpage


\begin{thebibliography}{}
\bibitem{footnote1}
We focus on one crossing in this study. Having more crossings may require fine tuning of the models or it may easily fall into the unstable branch of compact stars.
\bibitem{footnoteThermo}
The thermodynamics relation is $\mu(P)\approx \mu(0)+\int_0^P dP'\frac{1}{n(P')}\,$, where the number density $n(P)$ is positively correlated with the equation of states $\rho(P)$.
\end{thebibliography}
\end{document}